\documentclass{camera}
\usepackage{epsfig}
\begin{document}
\title{
	Measurement and analysis of neutron capture reaction rates of light 
	neutron--rich nuclei
}
\author{
	P.~Mohr$^{1}$, H.~Beer$^{2}$, H.~Herndl$^{1}$, H.~Oberhummer$^{1}$
}
\organization{
	$^1$ Institut f\"ur Kernphysik, Technische Universit\"at Wien, \\
	Wiedner Hauptstra{\ss}e 8--10, A--1040 Vienna, Austria \\
	$^2$ Forschungszentrum Karlsruhe, Institut f\"ur Kernphysik III, \\
	P.O. Box 3640, D--76021 Karlsruhe, Germany
}
\maketitle
\begin{abstract}
Several neutron capture cross sections of light neutron--rich nuclei
were measured in the astrophysically relevant energy region of 5 to
200 keV. The experimental data are compared to calculations using the
direct capture model. The results are used for the calculation of neutron 
capture cross sections of unstable isotopes.
Furthermore, neutron sources with energies below
$E_{\rm{n}} \approx 10~{\rm{keV}}$ are discussed.
\end{abstract}
\section{Introduction}
\label{sec:intro}
Contrary to charged--particle reactions neutron--capture reactions can
contribute to the nucleosynthesis of light and heavy elements. For the
heavy elements (n,$\gamma$)--reactions are the basis of the s-- and r--process.
In the s--process nucleosynthesis of heavy elements neutron--capture
is dominated by the compound--capture mechanism. However, in the r--process
scenarios, in the $\alpha$--rich freeze--out, and in the s--process for 
light nuclei the direct--capture (DC) is important.

Presently 
the direct capture mechanism can be investigated experimentally
only for light nuclei near the border of beta stability.
For heavy nuclei direct capture is non--negligible only for nuclei
near magic numbers. We have investigated neutron--capture reactions
for many neutron--rich target nuclei in the mass region $A = 10-50$
(see Refs.~\cite{allerlei}).
The measurements
were carried out in the thermonuclear energy region at the Karlsruhe 
3.75 MV Van-de-Graaff
accelerator. Neutrons were produced with the $^7$Li(p,n)--reaction
near the reaction threshold giving a quasi--Maxwellian energy spectrum
with $kT = 25~{\rm{keV}}$ and with thin Li--targets at higher
energies. The cross sections were determined with the method
of fast cyclic activation technique \cite{beer94}. Additionally,
experiments at thermal energies were 
carried out at the reactor BR1, Mol, Belgium.
In this case also the primary transitions 
and their branching could be measured.
\section{Present Experiments}
\label{sec:exp}
\begin{figure}[h]
\begin{minipage}[h]{9.0cm}
\begin{center}
\epsfig{file=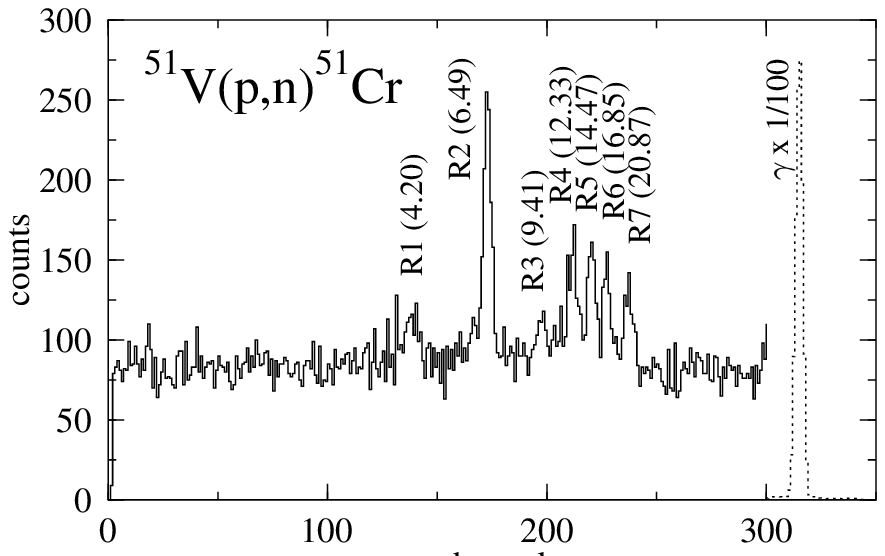,clip=off}
\end{center}
\caption{
	\label{fig:TOF}
	Neutron TOF spectrum of the reaction 
	$^{51}$V(p,n)$^{51}$Cr
	at $E_{\rm{p}} = 1592~{\rm{keV}}$,
	$\vartheta_{\rm{lab}} = 0^o$.
	For the known resonances we agree with
	Ref.~\protect\cite{kneff70}, and a new weak resonance
	was found ($E_{\rm{n}} = 4.20~{\rm{keV}}$).
}
\end{minipage}
\begin{minipage}[h]{0.4cm}
\begin{center}
\begin{tabular}{c}
\end{tabular}
\end{center}
\end{minipage}
\begin{minipage}[h]{6.0cm}
\begin{center}
\epsfig{file=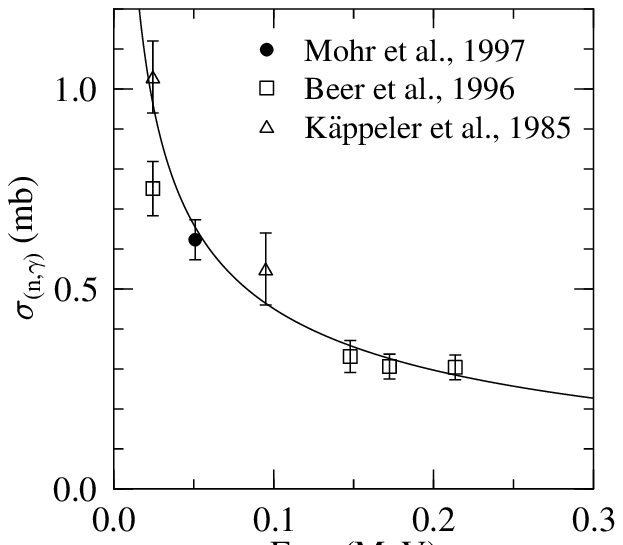,clip=off}
\end{center}
\caption{
	\label{fig:ca48}
	Experimental neutron capture cross section of $^{48}$Ca compared
	to a calculation using the DC model (see Ref.~\protect\cite{beer96}).
}
\end{minipage}
\end{figure}

In this work we extend our studies to neutron energies which cannot be
produced by the usual $^7$Li(p,n)--reaction. Because of the conservation of
momentum lower neutron energies than about 20 keV can only be obtained for
target nuclei with negative Q--values for the (p,n)--reaction which are
much heavier than $^7$Li. We analyzed the (p,n)--reactions of
$^{65}$Cu, $^{51}$V, and $^{45}$Sc using the pulsed proton beam and
a flight path of 91 cm between the production target and the neutron detector.
The neutrons are produced by the
resonances close above the (p,n) threshold. For $^{45}$Sc we confirmed
the result of Ref.~\cite{braunschweig} that one can obtain
practically monoenergetic neutrons at $E_{\rm{n}} = 8.15~{\rm{keV}}$
at $\vartheta_{\rm{lab}} = 0^o$.
For $^{51}$V we found a new resonance at $E_{\rm{n}} = 4.20~{\rm{keV}}$
(see Fig.~\ref{fig:TOF}) which is too weak to be used as neutron source
but strong enough to disturb the monoenergetic spectrum from the
$E_{\rm{n}} = 6.49~{\rm{keV}}$ resonance. For $^{65}$Cu 
the neutron yield is relatively small.

Additionally, a quasi-Maxwellian energy spectrum at $kT = 52~{\rm{keV}}$
can be obtained from the $^3$H(p,n)--reaction \cite{kae87}. Recently we 
used this spectrum to complete our measurement of the neutron capture 
cross section of $^{48}$Ca (see Fig.~\ref{fig:ca48}) \cite{mohr97}.
\section{Direct Capture Calculations, Reaction Rates}
\label{sec:DC}
The theoretical analysis was carried out in the framework of the
DC formalism \cite{kim87}. The capture cross section 
for each final state is determined
by the overlap of the scattering wave function $\chi_{l}(r)$, the 
electromagnetic transition operators $O^{E{\cal L},M{\cal L}}$
and the bound state wave function $u_{NL}(r)$, and by the spectroscopic
factor $C^2 S$ of the final state. The total capture cross section is given by
the sum of all partial capture cross sections:
\begin{equation}
\sigma^{DC} \sim \sum_i (C^2 S)_i \cdot 
	\Big| \int \chi_{l}(r) \; O^{E{\cal L},M{\cal L}} \;
	u_{NL,i}(r) \; d^3r \Big|^2
\label{eq:DC}
\end{equation}
For the calculation of the relevant wave functions we used folding potentials:
%
$V(R) =
  \lambda\,V_{\rm F}(R)
  =
  \lambda\,\int\int \rho_a({\bf r}_1)\rho_A({\bf r}_2)\,
  v_{\rm eff}\,(E,\rho_a,\rho_A,s)\,{\rm d}{\bf r}_1{\rm d}{\bf r}_2
\label{eq:fold}
%
$.
For the nucleus $^{48}$Ca the depth of the potentials 
was adjusted to the scattering length
(scattering wave) and to the binding energy (bound states)
leading to $\lambda$ values very close to unity \cite{beer96}.
The spectroscopic factors can be derived from Eq.~\ref{eq:DC}
with very small uncertainties
by comparing the calculated and the experimental capture cross section
at thermal energies \cite{beer96}.

For the calculation of the neutron capture cross sections of the unstable
Ca isotope $^{50}$Ca we can use the results derived for $^{48}$Ca. The
density distribution for $^{50}$Ca is obtained from a Hartree Fock
calculation. The strength parameter $\lambda$ of the optical
potential in the entrance channel is calculated using the same volume
integral as for $^{48}$Ca. The spectroscopic factors
of the isotopes $^{49,51}$Ca are calculated 
from the shell model using the code OXBASH \cite{bro84} and the
interaction FPD6 \cite{ric91}. 
The excitation energy of the $1/2^-$ state of $^{51}$Ca is 
$E_x = 1.773~{\rm{MeV}}$ using this interaction.

In Tab.~\ref{tab:tab1} we list spectroscopic factors of $^{51}$Ca
calculated from the above procedure, and we give 
reaction rate factors $N_A < \sigma \, v >$ calculated from the DC model.
The calcu\-lation of the neutron capture cross section of $^{50}$Ca 
gives somewhat smaller results than in a previous paper \cite{krausmann}.
However, the conclusions given in Ref.~\cite{krausmann}
remain unchanged.

\begin{table}[ht]
\caption{\label{tab:tab1} 
	Spectroscopic factors $C^2 S$ of $^{51}$Ca,
	and capture cross section $\sigma (30~{\rm{keV}})$
	and reaction rate factor $N_A < \sigma \, v >$ 
	of the reaction $^{50}$Ca(n,$\gamma$)$^{51}$Ca
}
\begin{center}
\begin{tabular}{cccccc}
$J^{\pi}$	& $E_x$ (MeV)	& $Q$ (MeV)	& $C^2 S$
	& $\sigma (30~{\rm{keV}}) \, ({\rm{mb}})$
	& $N_A < \sigma \, v > ({\rm{cm^3 s^{-1} mole^{-1}}})$	\\
\hline
\hline
$3/2^-$		& 0.0		& 4.400		& 0.47
	& 0.43
	& 6.20 $\times$ $10^4$ \\
$1/2^-$		& 1.773		& 2.627	& 0.82 
	& 0.18
	& 2.60 $\times$ $10^4$ \\
\hline
sum		&		&	&
	& 0.61
	& 8.80 $\times$ $10^4$ \\
\hline
\hline
\end{tabular}
\end{center}
\end{table}

{\footnotesize{
{\bf{\noindent{Acknowledgments:}}}
This work was supported by Fonds zur F\"orderung der Wissenschaftlichen
For\-schung (S7307--AST), Deutsche Forschungsgemeinschaft (Mo739), 
and Volkswagen--Stiftung (I/72286).
}

\end{document}